\documentclass[twocolumn,journal]{IEEEtran}
\usepackage[T1]{fontenc}
\usepackage[latin9]{luainputenc}
\usepackage{array}
\usepackage{amsmath}
\usepackage{graphicx}
\usepackage[unicode=true,
 bookmarks=true,bookmarksnumbered=true,bookmarksopen=true,bookmarksopenlevel=1,
 breaklinks=false,pdfborder={0 0 0},pdfborderstyle={},backref=false,colorlinks=false]
 {hyperref}
\hypersetup{pdftitle={Orthogonal Transform Multiplexing with Memoryless Nonlinearity: a Possible Alternative to Traditional Coded-Modulation Schemes},
 pdfauthor={Sergey V. Zhidkov},
 pdfpagelayout=OneColumn, pdfnewwindow=true, pdfstartview=XYZ, plainpages=false}

\makeatletter

\providecommand{\tabularnewline}{\\}

 \let\oldforeign@language\foreign@language
 \DeclareRobustCommand{\foreign@language}[1]{%
   \lowercase{\oldforeign@language{#1}}}

\usepackage[caption=false,font=footnotesize]{subfig}
\usepackage{balance}
\usepackage{cuted}
\makeatother

\begin{document}

\title{Orthogonal Transform Multiplexing with Memoryless Nonlinearity: a
Possible Alternative to Traditional Coded-Modulation Schemes }

\author{Sergey V. Zhidkov,~\IEEEmembership{Member,~IEEE}\thanks{S. V. Zhidkov is with Cifrasoft Ltd., Izhevsk, Russia, e-mail: \protect\href{mailto:sergey.zhidkov@cifrasoft.com}{sergey.zhidkov@cifrasoft.com}.}}

\markboth{Orthogonal Transform Multiplexing with Memoryless Nonlinearity}{ S. V. Zhidkov: Orthogonal Transform Multiplexing with Memoryless
Nonlinearity}
\maketitle
\begin{abstract}
In this paper, we propose a novel joint coding-modulation technique
based on serial concatenation of orthogonal linear transform, such
as discrete Fourier transform (DFT) or Walsh-Hadamard transform (WHT),
with memoryless nonlinearity. We demonstrate that such a simple signal
construction may exhibit properties of a random code ensemble, as
a result approaching channel capacity. Our computer simulations confirm
that if the decoder relies on a modified approximate message passing
algorithm, the proposed modulation technique exhibits performance
on par with state-of-the-art coded modulation schemes that use capacity-approaching
component codes. The proposed modulation scheme could be used directly
or as a pre-coder for a conventional orthogonal frequency division
multiplexing (OFDM) transmitter, resulting in a system possessing
all benefits of OFDM along with reduced peak-to-average power ratio
(PAPR).
\end{abstract}


\section{Introduction}

\IEEEPARstart{A}{} traditional approach for transmitting digital
information over a communication link with high spectral efficiency
($\eta\geq2$ bit/s/Hz) is to combine good error-correcting code with
high-order modulation. Over last two decades, several capacity-approaching
coded modulation techniques have been proposed, including low-density
parity check (LDPC) coded modulation {[}1, 2{]}, turbo-coded modulation
and turbo trellis-coded modulation {[}3{]}. In this paper, we propose
an alternative bandwidth-efficient modulation technique, which does
not rely on any error correction coding in the traditional sense,
while still achieving error rates close to the best known coded modulation
schemes. Our method is based on serial concatenation of an orthogonal
linear transform, such as DFT or WHT, with memoryless nonlinearity.
We demonstrate that such a simple signal construction may exhibit
properties of random code ensemble, as a result approaching channel
capacity if decoded using a maximum-likelihood (ML) decoder. 

When the proposed modulation scheme relies on DFT transform, it can
be viewed as an OFDM signal distorted by very strong memoryless nonlinearity.
In the context of OFDM, any nonlinearity is usually considered a highly
undesirable phenomenon. In fact, the sensitivity of OFDM technology
to nonlinear distortions is usually cited as a major drawback of OFDM.
Nonetheless, in {[}4{]}, the authors theoretically proved that the
bit error rate (BER) performance of OFDM transmission subjected to
strong memoryless nonlinearity can outperform linear transmission,
provided that an optimal ML-receiver is used. Moreover, in {[}5{]},
we proposed a practical message-passing receiver for nonlinearly distorted
OFDM signals, showing that the BER performance of hard-clipped OFDM
signals can be up to 2 dB better than linear OFDM transmission. In
this paper, we go one step further to demonstrate that if the uncoded
OFDM signal is distorted by memoryless nonlinearity with certain properties,
the resultant signal waveform may achieve BER performance close to
the best known capacity\textendash approaching coded modulation techniques.
The same performance can also be achieved by replacing DFT in the
transmitter with WHT, resulting in simpler hardware/software implementation
of the encoder and the decoder.

\section{Proposed modulation technique}

\subsection{Achieving capacity by serial concatenation of orthogonal transform
and memoryless nonlinearity}

\begin{figure}[tbh]
\includegraphics[scale=1.1]{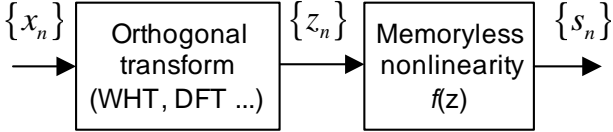}\caption{\label{fig:direct_modulator}Proposed modulation scheme}
\end{figure}

Consider the modulation scheme illustrated in Figure \ref{fig:direct_modulator}.
In the transmitter, $N$ uncoded real or complex baseband modulation
symbols $\ensuremath{\boldsymbol{\mathbf{x}}=\left\{ x_{k}\right\} }$
(e.g. $M$-QAM, or $M$-PAM) are first transformed by means of orthogonal
block transform and then passed through memoryless nonlinearity block.
The signal at the modulator output can be expressed as:

\begin{equation}
\ensuremath{s_{n}=f\left(z_{n}\right),\quad n=0,1,...,N-1}\label{eq: main_modulator}
\end{equation}
where $f(z)$ is a memoryless nonlinear function, and

\begin{equation}
\ensuremath{{\bf z}={\bf Fx},}
\end{equation}
where $\mathbf{z}=\left\{ z_{n}\right\} $, and $\mathbf{F}$ is a
$N\times N$ (unitary) orthogonal transform matrix. In this paper,
we consider two types of transforms: Walsh-Hadamard transform, and
discrete Fourier transform (real-valued or complex).

Let us now assume that the memoryless nonlinear function $f(z)$ can
be represented as $\ensuremath{f\left(z\right)=\underbrace{g\left(...g\left(g\left(z\right)\right)\right)}_{l-times}}$,
where $g(z)$ is a deterministic chaotic map. Thus, $f(z)$ has certain
properties of chaotic iterated functions, namely, \emph{sensitive
dependence on initial conditions} {[}6{]}. Under such assumptions,
and if $N\rightarrow\infty$ and $l\rightarrow\infty$, the ensemble
of waveforms generated by (\ref{eq: main_modulator}) possesses all
major properties of a random code ensemble, because even small modifications
in the message sequence $\ensuremath{\left\{ x_{k}\right\} }$ (e.g.,
in a single bit) lead to: 
\begin{description}
\item [{a)}] small modifications, at least, for all samples of the intermediate
signal $z_{n}$ due to the spreading properties of the orthogonal
transform operation, and 
\item [{b)}] large (and pseudo-random) modifications for all samples of
waveform $s_{n}$ due to the properties of the nonlinear function
$f(z)$. 
\end{description}
Therefore, we conjecture that where $f(z)$ has the above-mentioned
properties if the waveforms generated by (\ref{eq: main_modulator}) are
demodulated using a ML decoder, system performance may approach channel
capacity. Unfortunately, brute-force, maximum-likelihood decoding
of (\ref{eq: main_modulator}) is prohibitively complex, generally
of the order $O(M^{N})$, where $M$ is the modulation order. However,
as in many practical applications, we can rely on belief-propagation
techniques to approximate ML-decoding. A perfect candidate for decoding
the signals generated by (\ref{eq: main_modulator}) is the generalized
approximate message passing (GAMP) algorithm {[}7{]}. Having already
been applied to the decoding of clipped OFDM signals in {[}5{]}, it
has demonstrated good performance and reasonably good convergence
behavior. 

\subsection{GAMP algorithm and its modifications}

In the additive white Gaussian noise (AWGN) channel, the model of
the received signal can be expressed as

\begin{equation}
\ensuremath{y_{n}=f\left(z_{n}\right)+w_{n},}\quad n=0,1,...,N-1\label{eq:gamp_model}
\end{equation}
where $w_{n}$ is the AWGN term with zero mean and variance $\mu^{w}$.
Model (\ref{eq:gamp_model}) is equivalent to a general problem formulation
for the GAMP algorithm {[}7{]}, which belongs to a class of Gaussian
approximations of loopy belief propagation for dense graphs. In our
decoder implementation, we use the sum-product variant of the GAMP
algorithm, which approximates minimum mean-squared error estimates
of $\mathbf{x}$ and $\mathbf{z}$. 

During simulation study, we discovered that the convergence behavior
and bit and frame error rate (FER) performance of the conventional
GAMP algorithm applied to the decoding of (\ref{eq:gamp_model}) may
be improved by several simple algorithm modifications. In particular,
we used damped version of the GAMP algorithm, which is known to improve
convergence for certain mixing matrices {[}8{]}, and introduced scaling
factor for AWGN term variance $\mu^{w}\rightarrow\alpha\mu^{w}$.
In addition, we implemented the final decision selection and post-processing
scheme as will be explained in the next section that helped us overcome
certain deficiencies of a conventional GAMP algorithm. To reduce computational
complexity we opted to use a version of GAMP algorithm with scalar
step-sizes {[}7{]}. Note that the scalar step-size version of the
GAMP algorithm is optimal for WHT and complex DFT since in both cases
$\left|F_{n,k}\right|^{2}=1$. Moreover,  our experimental study revealed
that the performance loss for real DFT case ($\left|F_{n,k}\right|^{2}\neq1$)
was also marginal. 

The GAMP algorithm adapted to our problem is summarized below. The
algorithm generates a sequence of estimates ${\bf \hat{x}}\left(t\right)$,
${\bf \hat{z}}\left(t\right)$, for $t=1,2,...$ through the following
recursions:

\emph{Parameters:}

$t_{max}$ the maximum number of iterations, 

$\alpha$ the noise scaling factor, 

$\beta$ the damping factor.

\emph{Step 1) Initialization:}

$t=1$, ${\bf \hat{x}}\left(1\right)=\boldsymbol{0}$, $\tilde{\mathbf{x}}\left(1\right)=\boldsymbol{0}$,
${\bf \boldsymbol{\mu}}^{x}\left(1\right)=\boldsymbol{1}$, ${\bf \hat{s}}\left(0\right)=\boldsymbol{0}$

\emph{Step 2) Estimation of output nodes:}

\begin{equation}
\mu_{n}^{p}\left(t\right)=\frac{1}{N}\sum\limits _{k=0}^{N-1}\mu_{k}^{x}\left(t\right),\forall n\label{eq:gamp_algo_start}
\end{equation}

\begin{equation}
\hat{p}_{n}\left(t\right)=\sum\limits _{k=0}^{N-1}F_{n,k}\hat{x}_{k}\left(t\right)-\mu_{n}^{p}\left(t\right)\hat{s}_{n}\left(t-1\right),\forall n
\end{equation}

\begin{equation}
\hat{z}_{n}\left(t\right)=\frac{1}{C}\int\limits _{-\infty}^{\infty}ze^{-\frac{\left(y_{n}-f\left(z\right)\right)^{2}}{2\alpha\mu^{w}}-\frac{\left(\hat{p}_{n}\left(t\right)-z\right)^{2}}{2\mu_{n}^{p}\left(t\right)}}dz,\forall n\label{eq:Integral_start}
\end{equation}

\begin{equation}
\mu_{n}^{z}\left(t\right)=\frac{1}{C}\int\limits _{-\infty}^{\infty}z^{2}e^{-\frac{\left(y_{n}-f\left(z\right)\right)^{2}}{2\alpha\mu^{w}}-\frac{\left(\hat{p}_{n}\left(t\right)-z\right)^{2}}{2\mu_{n}^{p}\left(t\right)}}dz-\left(\hat{z}_{n}\left(t\right)\right)^{2},\forall n
\end{equation}
where 

\begin{equation}
C=\int\limits _{-\infty}^{\infty}e^{-\frac{\left(y_{n}-f\left(z\right)\right)^{2}}{2\alpha\mu^{w}}-\frac{\left(\hat{p}_{n}\left(t\right)-z\right)^{2}}{2\mu_{n}^{p}\left(t\right)}}dz,\forall n\label{eq:Integral_end}
\end{equation}

\begin{equation}
\hat{s}_{n}\left(t\right)=\left(1-\beta\right)\hat{s}_{n}\left(t-1\right)+\beta\dfrac{\hat{z}_{n}\left(t\right)-\hat{p}_{n}\left(t\right)}{\mu_{n}^{p}\left(t\right)},\forall n
\end{equation}

\begin{equation}
\mu_{n}^{s}\left(t\right)=\left(1-\beta\right)\mu_{n}^{s}\left(t-1\right)+\beta\frac{1-\frac{\mu_{n}^{z}\left(t\right)}{\mu_{n}^{p}\left(t\right)}}{\mu_{n}^{p}\left(t\right)},\forall n
\end{equation}

\emph{Step 3) Estimation of input nodes}: 

\begin{equation}
\tilde{x}_{k}\left(t\right)=\left(1-\beta\right)\tilde{x}_{k}\left(t-1\right)-\beta\hat{x}_{k}\left(t\right)
\end{equation}

\begin{equation}
\mu_{k}^{r}\left(t\right)=\left(\frac{1}{N}\sum\limits _{n=0}^{N-1}\mu_{n}^{s}\left(t\right)\right)^{-1},\forall k
\end{equation}

\begin{equation}
\hat{r}_{k}\left(t\right)=\tilde{x}_{k}\left(t\right)+\mu_{k}^{r}\left(t\right)\sum\limits _{n=0}^{N-1}F_{n,k}^{*}\hat{s}_{k}\left(t\right),\forall k
\end{equation}

\begin{equation}
\hat{x}_{k}\left(t+1\right)=\sum\limits _{m=1}^{M}d_{m}P_{m,k},\forall k
\end{equation}

\begin{equation}
\mu_{k}^{x}\left(t+1\right)=\sum\limits _{m=1}^{M}\left(d_{m}-\hat{x}_{k}\left(t+1\right)\right)^{2}P_{m,k},\forall k\label{eq:gamp_algo_end}
\end{equation}
where

\begin{equation}
P_{m,k}=\frac{e^{-\frac{\left(d_{m}-\hat{r}_{k}\left(t\right)\right)^{2}}{2\mu_{k}^{r}\left(t\right)}}}{\sum\limits _{l=1}^{M}e^{-\frac{\left(d_{l}-\hat{r}_{k}\left(t\right)\right)^{2}}{2\mu_{k}^{r}\left(t\right)}}},
\end{equation}
$M$ is the number of points in the signal constellation, and $\left\{ d_{m}\right\} $
is the vector of constellation points. For example, for 2-PAM modulation,
$\left\{ d_{m}\right\} =\left[-1{\rm \quad+}1\right]$, and for 4-PAM
modulation $\left\{ d_{m}\right\} =\left[-3\mathord{\left/\vphantom{-3{\sqrt{5}}}\right.\kern -\nulldelimiterspace}\sqrt{5}\quad-1\mathord{\left/\vphantom{-1{\sqrt{5}}}\right.\kern -\nulldelimiterspace}\sqrt{5}{\rm \quad+}1\mathord{\left/\vphantom{1{\sqrt{5}}}\right.\kern -\nulldelimiterspace}\sqrt{5}{\rm \quad+}{\rm 3}\mathord{\left/\vphantom{{\rm 3}{\sqrt{5}}}\right.\kern -\nulldelimiterspace}\sqrt{5}\right]$. 

Steps (\ref{eq:gamp_algo_start})\textendash (\ref{eq:gamp_algo_end})
are repeated with $t\rightarrow t+1$ until $t_{max}$ iterations
have been performed. 

We presented here the GAMP algorithm version for the real-valued modulation
and real orthogonal transform (e.g. $M$-PAM and WHT or real-DFT).
Nonetheless, the extension to the complex case is straightforward.
Moreover, in case of Cartesian-type nonlinearity $f(z)$ the complexity
of the decoding algorithm with $N$ complex input symbols is essentially
the same as the complexity of the real-valued algorithm with $2N$
input symbols.

More details on the GAMP algorithm and its thorough analysis can be
found in {[}7{]}. 

\subsection{Choosing optimal nonlinearity $f(z)$}

\begin{figure}[tbh]
\includegraphics[scale=0.61]{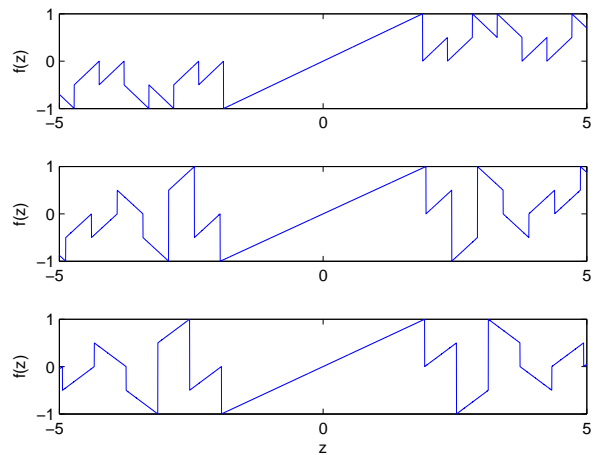}

\caption{\label{fig:Nonlinearities}Three examples of nonlinear functions $f(z)$:
nonlinearity 1 (top), nonlinearity 2 (middle), and nonlinearity 3
(bottom).}
\end{figure}

Choosing the optimal shape of nonlinearity $f(z)$ is not easy, requiring
a balance between two conflicting requirements. Firstly, the nonlinear
function should be reasonably \textquotedbl{}chaotic\textquotedbl{}
to guarantee sensitivity to initial conditions and, therefore, random-like
properties of the coded waveforms. On the other hand, our experimental
study implies that a \textquotedbl{}truly chaotic\textquotedbl{} shape
of nonlinearity $f(z)$ precludes GAMP algorithm from converging to
the ML-solution. Therefore, we adopted an \emph{ad hoc} procedure
to select and optimize the shape of the memoryless nonlinearity $f(z)$.
Our choice of nonlinearity $f(z)$ relies on two key ideas: 
\begin{itemize}
\item $f(z)$ should contain a linear or almost linear region around $f(0)$
to allow the message passing decoder to converge. 
\item $f(z)$ should \emph{resemble} a chaotic iterated function in other
regions to guarantee sensitivity to initial conditions. 
\end{itemize}
For a general representation of $f(z)$, we suggest using the flexible,
piece-wise linear model:

\begin{equation}
\ensuremath{f\left(z\right)=\left\{ \begin{array}{l}
{\mathop{\rm sgn}}\left(z\right)\left(a_{0}\left|z\right|+b_{0}\right),{\rm \quad if}\;0\le\left|z\right|<T_{1}\\
{\mathop{\rm sgn}}\left(z\right)\left(a_{1}\left|z\right|+b_{1}\right),{\rm \quad if\;}T_{1}\le\left|z\right|<T_{2}\\
...\\
{\mathop{\rm sgn}}\left(z\right)\left(a_{i-1}\left|z\right|+b_{i-1}\right),{\rm \quad if\;}T_{i-1}\le\left|z\right|<T_{i}\\
{\mathop{\rm sgn}}\left(z\right)\left(a_{i}\left|z\right|+b_{i}\right),{\rm \quad if\;}\left|z\right|\ge T_{i}
\end{array}\right.}\label{eq:Nonlinearity equation}
\end{equation}
with some parameters $\boldsymbol{a}=G_{0}\left\{ a_{0},a_{1},...,a_{i}\right\} $,
$\boldsymbol{b}=\left\{ b_{0},b_{1},...,b_{i}\right\} $, and $\boldsymbol{T}=G_{0}^{-1}\left\{ 0,T_{1},T_{2},...,T_{i}\right\} $,
where $G_{0}$ is a scaling factor that does not affect the ``shape''
of nonlinearity $f(z)$. For complex modulation (i.e. complex DFT),
we apply function (\ref{eq:Nonlinearity equation}) separately to
the real and imaginary components of $z$ (Cartesian-type nonlinearity).
To select the parameters of (\ref{eq:Nonlinearity equation}), we
devised dozens of different functions $f(z)$ that satisfy the above
mentioned heuristic requirements, and then optimized parameter $G_{0}$
at our target BER and FER (namely, BER$\leq10^{-5}$, FER$\leq$1\%).
This non-exhaustive search procedure yielded several ``good'' sets
of parameters $\boldsymbol{a}$, $\boldsymbol{b}$, and $\boldsymbol{T}$,
some of which are summarized in Table 1 and illustrated in Figure
\ref{fig:Nonlinearities}.

\begin{table}[tbh]
\caption{\label{tab:Nonlinearities}Example parameters of nonlinearity $f(z)$}

\centering{}%
\begin{tabular}{|c|>{\raggedright}p{0.85\columnwidth}|}
\hline 
\emph{No.} & \emph{Parameters}\tabularnewline
\hline 
1 & $\boldsymbol{a}G_{0}^{-1}$= \{1 2 2 -2 -2 2 2 -2 -2 -0.5\}\newline$\boldsymbol{b}$
= \{0 -2 -2.5 4 4.5 -4 -4.5 6 6.5 2.5\}\newline$\boldsymbol{T}G_{0}$=
\{0 1 1.25 1.5 1.75 2 2.25 2.5 2.75 3\}; $G_{0}=0.53$\tabularnewline
\hline 
2 & $\boldsymbol{a}G_{0}^{-1}$= \{1 2 2 -2 -2 2 2 -2 -2 -0.5\}\newline$\boldsymbol{b}$
= \{0 -2 -3.5 4 3.5 -4 -4.5 6 6.5 2.5\}\newline$\boldsymbol{T}G_{0}$=
\{0 1 1.25 1.5 1.75 2 2.25 2.5 2.75 3\}; $G_{0}=0.5125$\tabularnewline
\hline 
3 & $\boldsymbol{a}G_{0}^{-1}$= \{1.25 2 2 -2 -2 2 2 -2 -2 -0.5\}\newline$\boldsymbol{b}$
= \{0 -1.6 -3.1 3.6 3.1 -3.6 -4.1 5.6 6.1 2.4\}\newline$\boldsymbol{T}G_{0}$=
\{0 0.8 1.05 1.3 1.55 1.8 2.05 2.3 2.55 2.8\}; $G_{0}=0.415$\tabularnewline
\hline 
\end{tabular}
\end{table}

\section{Simulation results}

\begin{figure}[tbh]
\includegraphics[scale=0.53]{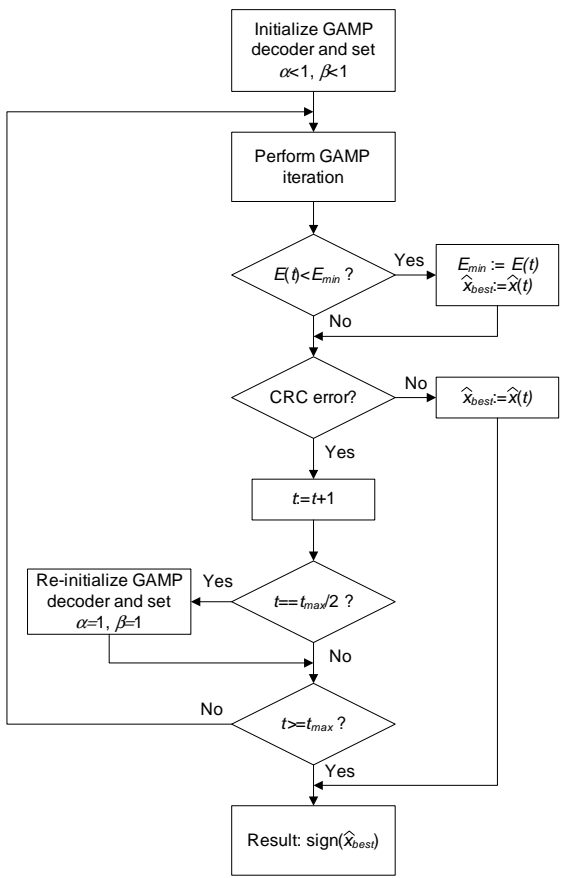}\caption{\label{fig:Decoding-algorithm-flowchart}Decoding algorithm flowchart}
\end{figure}

The performance of the proposed modulation scheme with the GAMP-based
decoder was studied by means of Monte-Carlo simulation. Our initial
simulation-based study revealed that the conventional GAMP algorithm
(with $\alpha=1$, and $\beta=1$) exhibits several deficiencies,
in particular, relatively high error-floor, slow convergence and occasional
instability problems (divergence with the increase of number of iterations).
It turns out that the error-floor issue was, at least, in part due
to the decoder algorithm. To alleviate these deficiencies we implemented
several enhancement techniques. Firstly, we appended the cyclic-redundancy
check (CRC) block to each data frame and used it for the early stopping
criterion. Secondly, after each decoding iteration we calculated the
Euclidean distance $E(t)$ between the received vector $\{y_{n}\}$
and the reconstructed waveform $f\left(\sum\limits _{k=0}^{N-1}F_{n,k}\hat{x}_{k}\left(t\right)\right)$,
and if a CRC error was detected after $t_{max}$ iterations the final
decision was based on the vector $\left\{ \hat{x}_{k}\left(t\right)\right\} $
that corresponded to the minimum Euclidean distance $E(t)$. Although
this procedure did not improve FER, it minimized the BER for incorrectly
decoded frames. Thirdly, we experimentally discovered that the decoder
convergence speed can be significantly improved by using noise scaling
factor $\alpha<1$ simultaneously with damping ($\beta<1$). However,
such a technique could result in occasional algorithm divergence,
therefore to  achieve better performance we implemented the following
procedure: $t_{max}/2$ iterations were performed using the modified
GAMP algorithm ($\alpha<1$, $\beta<1$), and if the decoder could
not converge within the first $t_{max}/2$ iterations, all internal
variables were reset and the subsequent $t_{max}/2$ iterations were
performed with the conventional GAMP settings ($\alpha=1$, $\beta=1$).
The overall decoding algorithm flowchart is depicted in Figure \ref{fig:Decoding-algorithm-flowchart}.

Our simulation results indicate that the performance of the proposed
scheme with WHT, real DFT or complex DFT with Cartesian-type nonlinearity
is almost identical. Therefore, from complexity point of view it seems
preferable to use WHT, and, hence, most of our simulation results
reported here were obtained for WHT-based scheme.

\begin{figure}[tbh]
\includegraphics[scale=0.57]{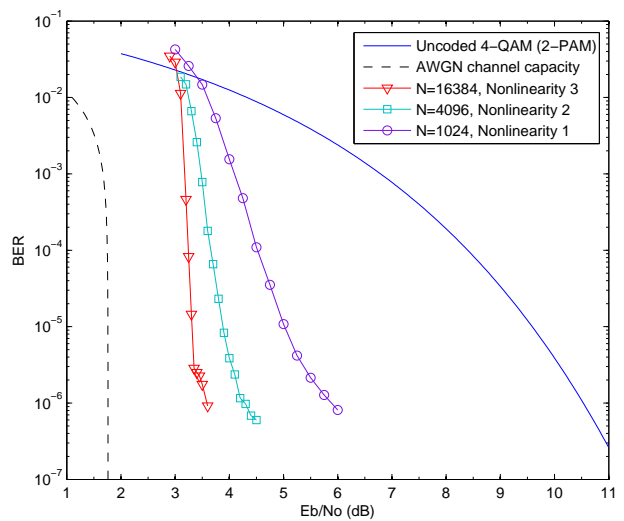}\caption{\label{fig:BER-vs-EbNo-capacity}BER vs $E_{b}/N_{0}$ for the proposed
modulation scheme with different nonlinearities and frame sizes in
AWGN channel ($\eta=2$ bit/s/Hz)}
\end{figure}

Figure  \ref{fig:BER-vs-EbNo-capacity} illustrates the BER vs $E_{b}/N_{0}$
curves for  2-PAM input modulation with three nonlinearities (Table
\ref{tab:Nonlinearities}/Figure \ref{fig:Nonlinearities}) and different
values of $N$. In all simulations, the variance of $\{z_{n}\}$ was
normalized to 1, the maximum number of iterations ($t_{max}$) was
set to 100, and integrals (\ref{eq:Integral_start})\textendash (\ref{eq:Integral_end})
were approximated using numerical summation. At the initialization
step the parameters $\alpha$, $\beta$ were set to $\alpha=0.71$,
$\beta=0.875$. 

\begin{figure}[tbh]
\includegraphics[scale=0.46]{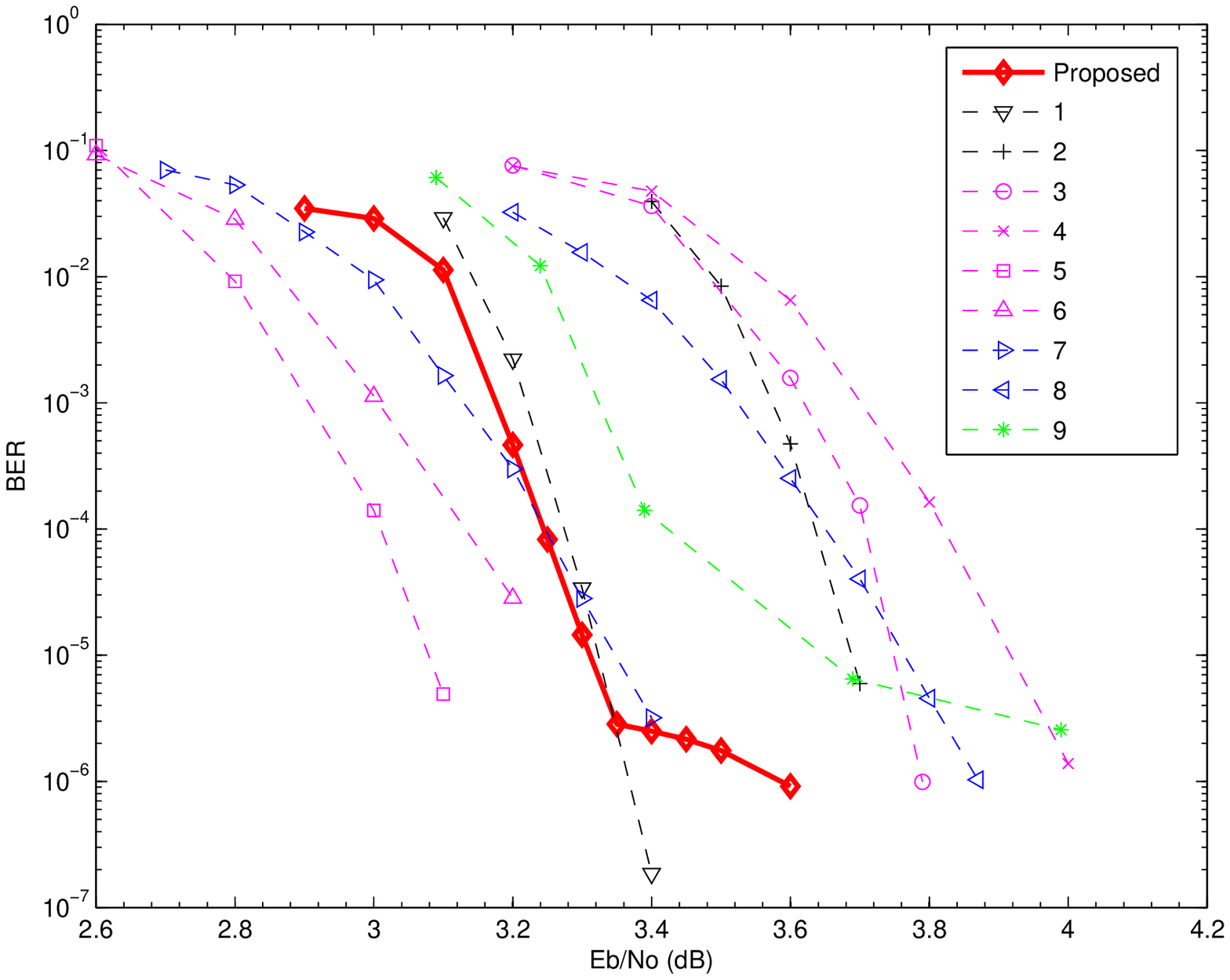}

\caption{\label{fig:BER-vs-EbNo}Performance comparison of the proposed technique
(WHT, Nonlinearity 3, $N$=16384) with state-of-the-art coded modulation
schemes ($2$ bit/s/Hz):\protect \\
1) ARJ4A LDPC, $N$=16384, 16-APSK {[}9{]}, \protect \\
2) ARJ4A LDPC, $N$=16384, 8-PSK {[}9{]}, \protect \\
3) Regular LDPC, BICM, $d_{v}$=3, $n$=20000, 4-PAM {[}1{]}\protect \\
4) Regular LDPC, BICM, $d_{v}$=3, $n$=10000, 4-PAM {[}1{]}\protect \\
5) Irregular LDPC, BICM, $d_{v}$=15, $n$=20000, 4-PAM {[}1{]}\protect \\
6) Irregular LDPC, BICM, $d_{v}$=15, $n$=10000, 4-PAM {[}1{]}\protect \\
7) eIRA LDPC, BICM, $n$=10000, 4-PAM {[}10{]}\protect \\
8) eIRA LDPC, BICM, $n$=9000, 8-PSK {[}10{]}\protect \\
9) Turbo TCM, $N$=10000, 8-PSK {[}3{]}}
\end{figure}

Remarkably, for the selected nonlinearity parameters $G_{0}$, $\boldsymbol{a}$,
$\boldsymbol{b}$, $\mathbf{T}$ the proposed modulation scheme exhibits
behavior similar to random-like codes with the presence of a waterfall
region and error-floor, and, as expected, performance improves for
larger frame sizes. Moreover, the proposed modulation scheme with
nonlinearity 3 and $N$=16384 achieves target BER=$10^{-5}$ at $E_{b}/N_{0}=3.3$
dB, which represents 6.3 dB gain over uncoded 2-PAM or 4-QAM modulation
and is only about 1.5 dB away from the \emph{unconstrained} AWGN channel
capacity. These results are better or within $0.1\div0.3$ dB of the
performance of the capacity-approaching, bandwidth-efficient modulation
schemes with $\eta=2$ bit/s/Hz reported in {[}1-3, 9, 10{]}. BER
curves for some of these advanced coded modulation schemes with comparable
frame sizes are reproduced in Figure \ref{fig:BER-vs-EbNo} for comparison. 

Although we set $t_{max}=100$, the average number of decoder iterations
was significantly lower. For example, at $BER=10^{-5}$ the average
number of decoder iterations was about 26 for the frame size $N$=16384,
and about 17 for the frame size $N$=4096, and only 7 iterations for
the frame size $N$=1024.

A considerable performance improvement over uncoded modulation was
also observed for 4-PAM or 16-QAM input modulation ($\eta=4$ bit/s/Hz).
However, in case of 4-PAM/16-QAM, the  GAMP algorithm is sub-optimal
in terms of BER performance, and the example nonlinearities that we
use for 4-QAM/2-PAM (Table \ref{tab:Nonlinearities}) are not optimal
either. Therefore, application of the proposed technique to high-order
modulation formats ($\eta\geq4$ bit/s/Hz) is an open research topic. 

\section{Discussion}

\begin{figure}[tbh]
\includegraphics[scale=0.8]{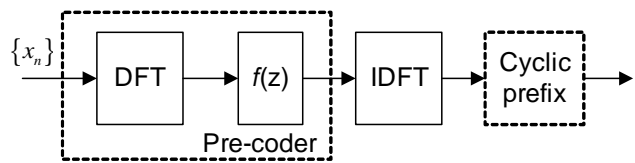}

\caption{\label{fig:precoder_for_ofdm}Proposed modulation scheme as a pre-coder
for a conventional OFDM transmitter}
\end{figure}

\begin{figure}[tbh]
\includegraphics[scale=0.5]{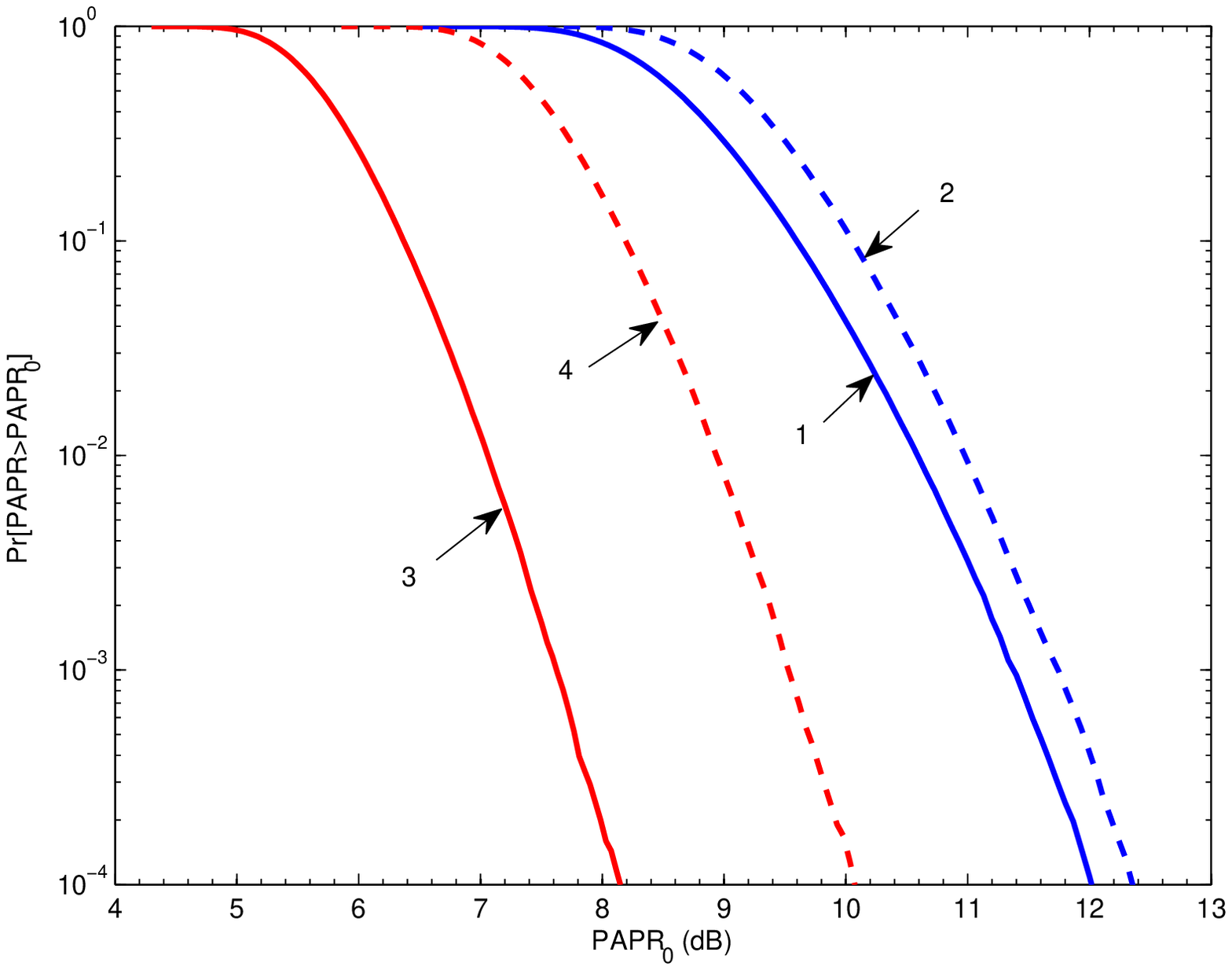}

\caption{\label{fig:CCDF}CCDF of PAPR for the proposed pre-coded OFDM (Figure
\ref{fig:precoder_for_ofdm}) with $N$=1024 and nonlinearity \#1:\protect \\
1) Conventional OFDM (Nyquist sampling)\protect \\
2) Conventional OFDM (four-times oversampling)\protect \\
3) OFDM with the proposed modulation (Nyquist sampling)\protect \\
4) OFDM with the proposed modulation (four-times oversampling)}
\end{figure}

We believe that the proposed joint coding-modulation technique might
be useful in some wired and wireless communication applications, since
it has several interesting properties:
\begin{itemize}
\item Good BER performance: As illustrated in the previous section the BER
performance is on par with that of the best known coded modulation
schemes.
\item Relatively low decoder complexity: The GAMP-based decoder complexity
per iteration is dominated by two orthogonal transform operations,
which, in case of fast WHT, require only $N\log_{2}(N)$ real additions
or subtractions. The input/output nonlinear step is generally a scalar
operation with complexity of order $O(N)$. Although, in our simulation
model, we used numerical integration to approximate integrals (\ref{eq:Integral_start})\textendash (\ref{eq:Integral_end}),
these can be expressed in closed-form using tabulated Gauss error
function, and, consequently, can be simplified or implemented via
lookup tables. It is also possible to use simpler max-sum version
of the GAMP algorithm to trade-off performance and complexity {[}7{]}. 
\item The choice of nonlinearity $f(z)$ may be tailored to the requirements
of the transmission system in order to improve overall system efficiency.
For example, $f(z)$ may be selected to improve power conversion efficiency,
or illumination-to-communication conversion efficiency in visible
light communication applications {[}11{]}.
\item An interesting application of the proposed modulation scheme is to
use it in combination with a conventional, linear OFDM transmitter,
as illustrated in Figure \ref{fig:precoder_for_ofdm}. Such a system
arrangement may result in the OFDM signal with reduced PAPR. Due to
the presence of nonlinearity $f(z)$, the distribution of signal samples
at the output of the modulator illustrated in Figure \ref{fig:precoder_for_ofdm}
resembles the distribution of $M$-QAM/$M$-PAM signal affected by
additive Gaussian noise. It can be shown that if the nonlinearity
$f(z)$ has a relatively large linear region the PAPR of such a signal
will be much lower than that of a conventional OFDM signal. Figure
\ref{fig:CCDF} compares the complementary cumulative distribution
functions (CCDF) of PAPR for signals generated by the transmission
system illustrated in Figure \ref{fig:precoder_for_ofdm} with nonlinearity
1 (complex-valued OFDM with Cartesian-type nonlinearity), and a conventional
OFDM system. We analyzed CCDF for Nyquist sampled waveforms and four-times
oversampled waveforms, since four-times oversampling provides a good
approximation of the continuous-time PAPR {[}12{]}. As may be seen,
at probability $10^{-4}$ the PAPR of the continuous signal generated
by the transmission system illustrated in Figure \ref{fig:precoder_for_ofdm}
is approximately 2.3 dB lower than the PAPR of a conventional OFDM
signal. The difference is increased to 3.9dB for Nyquist-sampled waveforms.
\end{itemize}
It should be noted that the proposed method has several limitations
and open issues. Firstly, in this paper, we focused primarily on the
case $\eta=2$ bit/s/Hz. One possible way to apply this technique
to systems with higher spectral efficiency is to use the output waveform
puncturing. This approach (i.e. random puncturing) seems to work well
up to $\eta=2.5\div3$ bit/s/Hz. It is also possible to use higher-order
input modulations (16-QAM/4-PAM), however, as we mentioned earlier,
extension of this technique to spectral efficiency $\eta\geq4$ bit/s/Hz
is still an open research topic. Secondly, the GAMP algorithm is based
on Gaussian and quadratic approximations, and, therefore, it is apparently
sub-optimal, especially, for small frame sizes. We were able to improve
performance of the GAMP-based decoder using several \emph{ad hoc}
techniques. However, it is still unclear, how close the GAMP decoder
performance is to the theoretical ML performance. Finally, our choice
of nonlinearity $f(z)$ is generally based on a heuristic and experimental
approach. The problem here is that the performance of the proposed
scheme is limited not only by distance distribution of encoded waveforms
but also by non-idealities of the decoding algorithm, and this fact
greatly complicates optimization strategy. 


\balance

\section{Conclusions}

In this paper, we proposed a novel joint coding-modulation technique
based on serial concatenation of orthogonal linear transformation
(e.g., WHT or DFT) with memoryless nonlinearity. We demonstrated that
such a simple signal construction may exhibit properties of a random
code ensemble, as a result approaching channel capacity. Our computer
simulations confirmed that if the decoder relies on the approximate
message passing algorithm, the proposed modulation technique exhibits
performance on par with state-of-the-art coded modulation schemes
that use capacity-approaching component codes. The proposed technique
could be extended to modulation formats with higher spectral efficiency
($\eta\geq4$ bit/s/Hz) and other types of orthogonal transformations,
offering one possible direction for our future research.

\end{document}